\documentclass[aps,prd,groupedaddress]{revtex4}

\usepackage{xcolor}
\usepackage[normalem]{ulem}

\usepackage{graphicx}
\usepackage{epsfig}
\usepackage{amsfonts}
\usepackage{amssymb}
\usepackage{epsf}
\usepackage{color}
\newcommand{\insertplot}[5]{\begin{figure}
 \hfill\hbox to 0.05in{\vbox to #5in{\vfill
 \inputplot{#1}{#4}{#5}}\hfill}
 \hfill\vspace{-.1in}
 \caption{#2}\label{#3}
 \end{figure}}
\newcommand{\inputplot}[3]{
 \special{ps: plotfile #1}
}

\newcounter{fig}

\newcommand{\ee}{\end{equation}}
\newcommand{\eea}{\end{eqnarray}}
\newcommand{\be}{\begin{equation}}
\newcommand{\bea}{\begin{eqnarray}}

\begin{document}

\title{Horndeski Proca stars with vector hair}
\author{Yves Brihaye}
\email[]{yves.brihaye@umons.ac.be}
\affiliation{Service de Physique de l’Univers, Champs et Gravitation, Universit\'e de Mons, Mons, Belgium}
\author{Betti Hartmann}
\email[]{b.hartmann@ucl.ac.uk}
\affiliation{Department of Mathematics, University College London, Gower Street, London, WC1E 6BT, UK}
\author{Burkhard Kleihaus}
\email[]{b.kleihaus@uni-oldenburg.de}
\affiliation{Institute of Physics, University of Oldenburg, D-26111 Oldenburg, Germany}
\author{Jutta Kunz}
\email[]{jutta.kunz@uni-oldenburg.de}
\affiliation{Institute of Physics, University of Oldenburg, D-26111 Oldenburg, Germany}

\date{\today}

\begin{abstract}
We study Proca stars in a vector-tensor gravity model inspired by Horndeski's generalized Einstein-Maxwell
field equations, supplemented with a mass term
for the vector field. We discuss the 
effects of the non-minimal coupling term on the 
properties of the resulting Proca stars. 
We show that the sign of the coupling constant is crucial
in determining the generic properties of these generalized
Proca star solutions, as soon as the magnitude of the
coupling constant is sufficiently large to allow for
significant deviations from the standard Proca star case.
For negative coupling constant we observe a new type of
limiting behavior for the generalized Proca stars,
where the spacetime splits into an interior region with matter fields and an exterior Schwarzschild region.

\end{abstract}

\maketitle

\section{Introduction}
In recent times, extended gravity models have received much interest (see, e.g., 
\cite{Will:2005va,Faraoni:2010pgm,Berti:2015itd,CANTATA:2021mgk}).
This big impetus resides on the one hand on cosmological issues they might resolve and on the other hand 
on the advent of gravitational wave multi-messenger
astronomy
\cite{Coulter:2017wya,LIGOScientific:2017vwq,LIGOScientific:2017ync,LIGOScientific:2018hze},
which is now allowing to test such models in new regimes. 
Amongst the models studied are numerous models containing
new degrees of freedom associated with additional scalar or vector type gravitational fields and non-minimal couplings between such scalar or vector and tensor fields. 

In fact, by requiring the equations of motion to remain second order, Horndeski proposed already in 1974 a general Lagrangian in \cite{Horndeski:1974wa,Charmousis:2011bf,Kobayashi:2011nu}  based on 
an extension of General Relativity by a real scalar field.
In 1976 Horndeski then extended Einstein-Maxwell theory
and obtained the most general second-order vector-tensor
theory of gravitation and electromagnetism, subject to
several conditions \cite{Horndeski:1976gi}.
When relaxing these conditions more general vector-tensor theories with second order field equations arise  \cite{Tasinato:2014eka,Heisenberg:2014rta}, 
similar to the case of Horndeski-type scalar-tensor theories
(see \cite{Nicosia:2020egv} for the teleparallel version of such generalized Proca theories).

A central point of interest in extended gravity models are certainly their black hole solutions. 
Unlike the case of scalar-tensor theories, however, vector-tensor theories have received much less attention in this respect.
Black hole solutions in the vector-tensor model 
of Horndeski \cite{Horndeski:1976gi}, that features
gauge invariance, were examined already in \cite{Muller:1988} and generalized recently in \cite{Verbin:2020fzk}. 
Black hole solutions of the more general vector-tensor theories \cite{Tasinato:2014eka,Heisenberg:2014rta} were discussed in \cite{Chagoya:2016aar,Babichev:2017rti,Chagoya:2017fyl,Heisenberg:2017xda,Heisenberg:2017hwb,Fan:2016jnz}.
Analogous to the phenomenon of spontaneous scalarization of black holes \cite{Antoniou:2017acq,Doneva:2017bvd,Silva:2017uqg},
recently also the phenomenon of spontaneous vectorization of black holes was argued to occur \cite{Ramazanoglu:2017xbl,Ramazanoglu:2018tig,Ramazanoglu:2019gbz,Ramazanoglu:2019jrr},
when the vector field is suitably coupled to an invariant.
Indeed, in \cite{Barton:2021wfj} it was shown that black holes can form Proca hair ``spontaneously'', when
the standard Einstein-Proca theory is extended to include a non-minimal coupling of the form $A_{\mu} A^{\mu}{\cal I}$ with ${\cal I}$ the Gauss-Bonnet invariant.
 
The presence of additional fields of scalar or vector type may, however, also lead to regular gravitating solutions.
In the case of complex fields (which are equivalent to doublets of real fields) these may constitute boson stars
as first conceived in 
\cite{Feinblum:1968nwc,Kaup:1968zz,Ruffini:1969qy}
for minimally coupled scalar fields.
Such boson stars with scalar fields have been widely investigated ever since (see e.g.,
\cite{Lee:1991ax,Jetzer:1991jr,Schunck:2003kk,Liebling:2012fv}).
But scalar boson stars have also been studied in a variety of extended gravity models
\cite{Whinnett:1999sc,Alcubierre:2010ea,Ruiz:2012jt,Hartmann:2013tca,Brihaye:2013zha,Kleihaus:2015iea,Brihaye:2016lin,Baibhav:2016fot,Brihaye:2018grv}.
In contrast, boson stars composed of vector fields, i.e., Proca stars, have only been considered in more recent times.

Proca stars were first obtained for the case of a  massive vector field coupled minimally to General Relativity in \cite{Brito:2015pxa}, where it was shown
that Proca stars are similar to scalar boson stars in many respects. 
They are globally regular solutions of the (extended) Einstein-Proca equations, where the harmonic time dependence (with frequency $\omega$) of the complex Proca field cancels out in the stress-energy tensor, and therefore leads to a stationary space-time.
The global U(1) invariance of the theory gives rise to a conserved Noether charge, the particle number.
Proca stars form a characteristic spiraling pattern,
when the mass or the charge are considered as functions of the boson frequency.
Subsequent studies considered Proca stars with self-interaction \cite{Brihaye:2017inn,Minamitsuji:2018kof} and charge \cite{SalazarLandea:2016bys}, their time evolution \cite{Sanchis-Gual:2017bhw}, formation \cite{DiGiovanni:2018bvo}, and collisions \cite{Sanchis-Gual:2018oui}, 
as well as Proca stars with non-minimal coupling to gravity \cite{Minamitsuji:2017pdr}.

The role of this paper is to emphasize different  -- and new -- aspects of  gravitating Proca stars, when extended gravity models are employed.
Along with \cite{Heisenberg:2017xda,Heisenberg:2017hwb,Barton:2021wfj} we here consider a vector-tensor model, but allow for a complex vector field. 
In particular, we employ the invariant of 
the generalized Horndeski Einstein-Maxwell theory
\cite{Horndeski:1976gi} as
the non-minimal vector-tensor coupling term,
supplemented by a mass term.
We construct the Proca stars of this extended gravity model
and investigate their properties.
This includes the dependence on the coupling constant
and the limiting behavior of the resulting families of
Proca stars.

This paper is organized as follows: 
We present the model, the ansatz and the
resulting field equations in section 2.
We discuss the generalized Proca stars
and their properties for positive and
negative coupling constant in section 3,
where we also recall the properties
of the standard Proca stars for comparison.
We conclude in section 4.

\section{The Model}

We consider the following non-minimally coupled vector-tensor model 
\be
\label{action}
   S = \int {\rm d}^4 x \sqrt{-g} \bigg[ 
   \frac{1}{2\kappa} R 
	- \frac{1}{4} F^*_{\mu \nu}F^{\mu \nu} 
	+ \gamma {\cal I}(g,A,A^*) 
	- U(A^*_{\mu}A^{\mu}) 
	\bigg]
\label{lagrangian}
\ee
where $R$ is the Ricci scalar and $F_{\mu \nu}$ is the field strength tensor of a complex vector field $A_\mu$. The
non-minimal coupling term between the vector field and the tensor field
${\cal I}(g,A,A^*)$ reduces for a real vector field
to the coupling term of the generalized Einstein–Maxwell theory of Horndeski \cite{Horndeski:1974wa,Horndeski:1976gi},
which uniquely satisfies the following conditions:
it yields second-order vector-tensor field equations via a variational principle, 
it yields charge conservation, and it yields the
Maxwell equations in the flat space limit.
The coupling term reads:
\be
\label{horndeski}
{\cal I}(g,A,A^*) = -\frac{1}{4} (F^*_{\mu \nu} F^{\kappa \lambda} R^{\mu \nu}_{\phantom{\mu \nu} \kappa \lambda}
                          - 4 F^*_{\mu \kappa} F^{\nu \kappa} R^{\mu}_{{\phantom \mu} \nu}
													+ F^*_{\mu \nu} F^{\mu \nu} R ) \ ,
\ee
and its strength is governed by the coupling constant $\gamma$.
Whereas Horndeski theory features gauge invariance
of the vector field,
we here break gauge invariance 
by adding a potential $U(\psi)$ for the vector field, 
\be
     U(\psi) = \frac{\mu^2}{2} \psi + \frac{\lambda}{4} \psi^2 \  , \ \ \ \psi = A^*_{\mu}A^{\mu} \  ,
\ee
that contains  
a mass term and a self-interaction term. 

The Lagrangian (\ref{action}) possesses a global $U(1)$ symmetry $A_{\mu} \to \exp(i \chi)A_{\mu}$ with an associated conserved 
Noether current of the form
\be
\label{current}
        j^{\alpha} = \frac{i}{2}((F^{\alpha \beta})^* A_{\beta} -F^{\alpha \beta} A^*_{\beta}) (1 + \gamma R) 
				+ \frac{i \gamma}{2}(A_{\beta} F^{* \mu \nu} - A^*_{\beta} F^{\mu \nu}) R^{\alpha \beta}_{\phantom{\alpha \beta} \mu \nu } 
				- 2 i \gamma((F^{\mu \beta})^* A_{\beta}-F^{\mu \beta} A^*_{\beta}) R_{\mu}^{\alpha} \ .
\ee

We will follow \cite{Horndeski:1976gi} to derive the Einstein and vector field equations. 
Variation of the action (\ref{action}) with respect to the metric $g_{ij}$ yields the generalized Einstein equations
\begin{equation}
\label{eq:einstein}
A^{ij} = 0 = 
\sqrt{-g}
\left\{-\frac{1}{2\kappa} G^{ij}-\frac{\gamma}{8} A_{({\cal I})}^{ij}   +\frac{1}{2} A_{(F)}^{ij}-A_{(U)}^{ij}\right\}
\end{equation}   
with 
\begin{eqnarray}
A_{({\cal I})}^{ij} & = &  \delta^{iabc}_{defk} g^{dj}\left\{
                           \frac{1}{2} \left(\left[F^{*}_{al} F^{el} +F_{al} F^{*el}\right] R_{bc}^{\phantom{bc}fk} 
		           +\nabla^k F^{*}_{al} \nabla_c F^{el}+\nabla^k F_{al} \nabla_c F^{*el}\right)
		             \right\}  \ , 
\nonumber\\
A_{(F)}^{ij} & = & \frac{1}{2} \left(F^{*ia} F^j_{\phantom{j}a} +F^{ia} F^{*j}_{\phantom{ia}a}\right)
                   -\frac{1}{4} g^{ij} \frac{1}{2} \left( F^{*ab}F_{ab}  + F^{ab} F^{*}_{ab} \right)  \ , 
\nonumber\\
A_{(U)}^{ij} & = & \frac{\partial U}{\partial \psi} \frac{\partial\psi}{\partial g_{ij}}
                +\frac{1}{2} U g^{ij}   \ , 
\nonumber		
\end{eqnarray}
and $\delta^{abcd}_{efkl} = \eta^{abcd}\eta_{efkl}$, where $\eta^{abcd}$ denotes the 
Levi-Civita tensor.

On the other hand, variation with respect to the vector field $A^*_i$ yields the vector field equations
\begin{equation}
\label{eq:proca_field}
B^i = 0 = 
-\frac{\gamma}{4}  \delta^{iabc}_{defk}\nabla_a F^{de}R_{bc}^{\phantom{bc}fk} 
+ \nabla_j F^{ij}
+2 \frac{\partial U}{\partial \psi} A^i \ .
\end{equation}   

\subsection{Ansatz}

We will be interested in spherically symmetric solutions and hence choose the metric to be of the form
\be
     {\rm d}s^2 = - f(r) (\sigma(r))^2   {\rm d}t^2 + \frac{1}{f(r)}  {\rm d}r^2 + r^2\left( {\rm d}\theta^2 +\sin^2 \theta {\rm d}\varphi^2  \right) \  , \ \ \ f(r)=1-\frac{2m(r)}{r} \ ,  
\ee
while the compatible most general Ansatz for the vector field reads
\be
 \ \ A_{\mu}  {\rm d}x^{\mu}=   e^{-i \omega t}
 \left[ a_0(r)  {\rm d}t  + i a_1(r)  {\rm d}r \right]    \ .
\ee
With this Ansatz, the non-vanishing components of the field strength and the non-minimal coupling term read, respectively:
\be
    F_{tr} = \omega a_1 - a_0' \  , \ \ \ {\cal I} =    \frac{(\omega a_1 - a_0')^2(1 - f)}{\sigma^2 r^2}  \ ,
\ee
where the prime denotes the derivative with respect to $r$.

The reduced  effective Lagrangian density then becomes
\be
 {\cal L}_{\rm eff}  =   
 \frac{1}{\kappa}\left(\sigma(1-f-r f') \right) 
 +   \frac{r^2}{2 \sigma} (\omega a_1 - a_0')^2 
  - \sigma r^2 U(\psi) 
 + \gamma \frac{(1-f)(\omega a_1 - a_0')^2}{\sigma} 
			\  , \ \ \ \psi \equiv 
				- \frac{a_0^2}{f \sigma^2 }
			+ f a_1^2  \ ,
\ee
yielding for the two metric functions $m(r)$ and $\sigma(r)$ the equations
\bea
\label{eq:m}
     m' = \frac{\kappa}{2} r^2 {T^0_0}_{\rm eff}
     \  \ \ \ ,  \  \ \ \
		 \sigma'=  \kappa \sigma r \left[\left(a_1^2 + \frac{a_0^2}{f^2 \sigma^2}\right) \left( \frac{\mu^2}{2} + \frac{\lambda}{2}\left(f a_1^2 - \frac{a_0^2}{f \sigma^2} \right)\right)
		         + \frac{\gamma}{r^2 \sigma^2}(\omega a_1- a_0')^2   \right]
\eea
with the effective energy density
\be
\label{eq:t00}
     \left({T^0_0}\right)_{\rm eff} = 
     \frac{(\omega a_1 - a_0')^2}{2 \sigma^2} \left(1 + \frac{2 \gamma (1-f)}{r^2}\right) + \frac{\mu^2}{2} \left(f a_1^2 + \frac {a_0^2}{f \sigma^2}\right)
		                                       + \frac{\lambda}{4} \left(f^2 a_1^4 + \frac{2}{\sigma^2} a_0^2 a_1^2 -\frac{3}{f^2 \sigma^4} a_0^4 \right)  \ .
\ee
The equations for the two Proca-field functions read
\be
\label{eq:proca1}
 \ \ \omega(a_0' - \omega a_1)\left[1 + \frac{2\gamma(1-f)}{r^2}\right] + 2 \sigma^2 f a_1  \frac{{\rm d} U}{{\rm d} \psi} = 0  \ 
\ee
and
\be
\label{eq:proca2}
 \ \  \frac{\sigma f}{r^2} \frac{d}{dr} \left[\frac{(a_0' - \omega a_1)}{\sigma}(r^2 + 2 \gamma(1-f))\right] - 2 a_0  \frac{{\rm d} U}{{\rm d} \psi}= 0 \ .
\ee
Combining the two  equations above, the Lorentz condition on the Proca field  can be obtained as follows (note that this condition is independent of the Horndeski term)
\be
     \nabla_{\mu} \left(\frac{{\rm d} U}{{\rm d} \psi}   A^{\mu} \right) = 0  \ \ \ \  \Longrightarrow  \ \ \ 
		\frac{\sigma f}{r^2} \frac{{\rm d}}{{\rm d}r} \Biggl( r^2 \sigma f a_1 \frac{{\rm d} U}{{\rm d} \psi}   \Biggl) + \omega a_0  \frac{{\rm d} U}{{\rm d} \psi}  = 0  \ .
\ee
We note that the equations (\ref{eq:m}) - (\ref{eq:proca2}) are consistent with the
general  Einstein and vector field equations (\ref{eq:einstein}) and (\ref{eq:proca_field}).

In order to solve the system of four coupled ordinary differential equations above, we need to impose appropriate boundary conditions in order to ensure regularity as well as asymptotic flatness. These conditions read
\be
      m(0) = 0 \ ,   
      \ \ \ a_1(0)=0 \  , \ \ \ a_0'(0) = 0 \  , \ \ \ \sigma(\infty)=1 \  , \ \ \ a_0(\infty) = 0  \ .
\ee 
There are five parameters, $\kappa$, $\gamma$, $\mu$, $\lambda$, and $\omega$. But instead of $\omega$ we use 
$a_0(0)$ as an input parameter, which we impose as an additional condition on the above system, supplemented with an auxiliary differential equation for $\omega$, $\omega'=0$.
Note that the parameters $\kappa$ and $\mu$ can be set to fixed values without losing generality by rescaling the vector field and the radial variable, suitably.
The remaining parameters are then $\gamma$, $\lambda$, and $\omega$ or $a_0(0)$.
Also note that $m(r\rightarrow\infty)\rightarrow M$, i.e. the value of the mass function $m(r)$ at spatial infinity gives the ADM mass $M$.

Within the spherical symmetric ansatz, the globally conserved charge associated with the locally conserved Noether current (\ref{current}) takes the form
\be
     Q = 4 \pi \int_0^{\infty} j^0 
     \sigma r^2 dr \  , \ \ \ j^0 
     = \frac{a_1(\omega a_1 - a_0')}{\sigma^2} \left[1 + \gamma\left(R + 2R^{01}_{\phantom{01}01} - 4 R^0_0\right)\right]
\ee
with
\be
R + 2R^{01}_{\phantom{01}01} - 4 R^0_0 = 2 \left(\frac{1-f}{r^2} + \frac{2f \sigma'}{r \sigma}\right)  \ . 
\ee

\section{Results}

\subsection{$\gamma=0$ case}

\begin{figure}[h!]
\begin{center}
\includegraphics[width=.45\textwidth, angle =0]{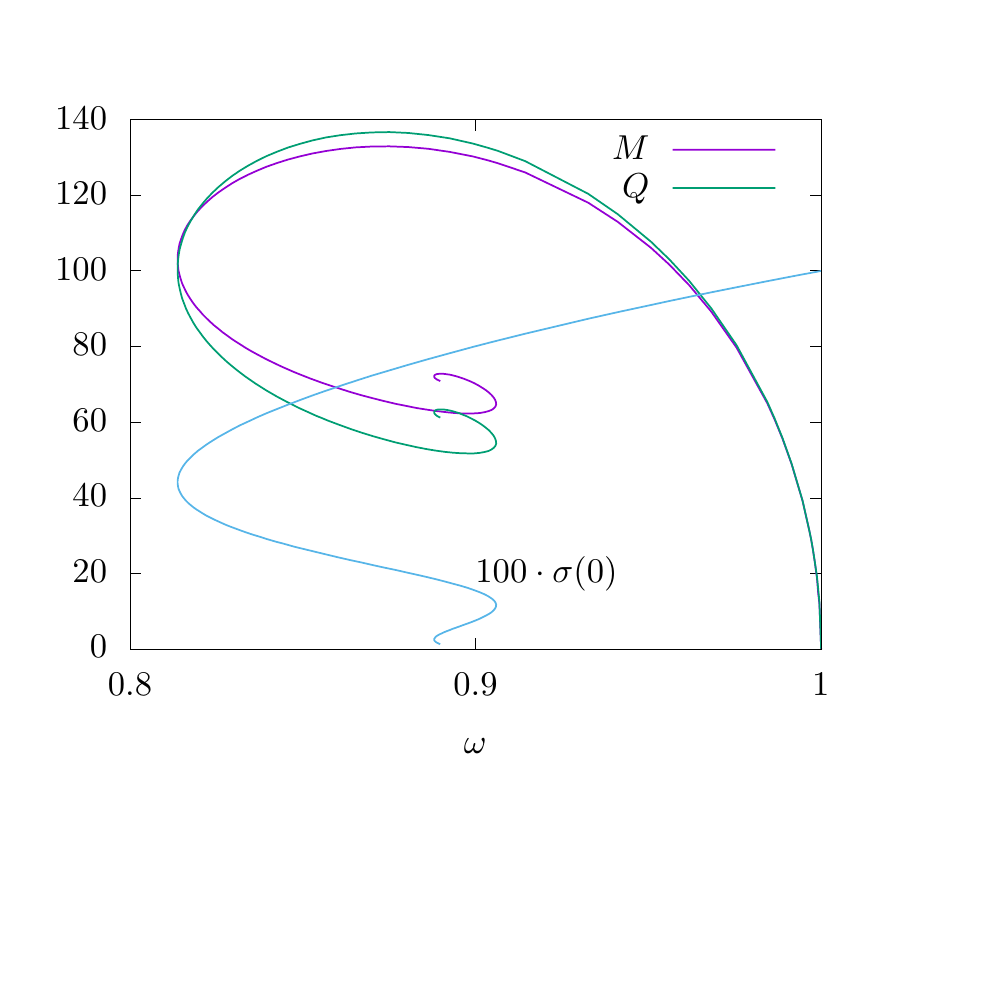}
\includegraphics[width=.45\textwidth, angle =0]{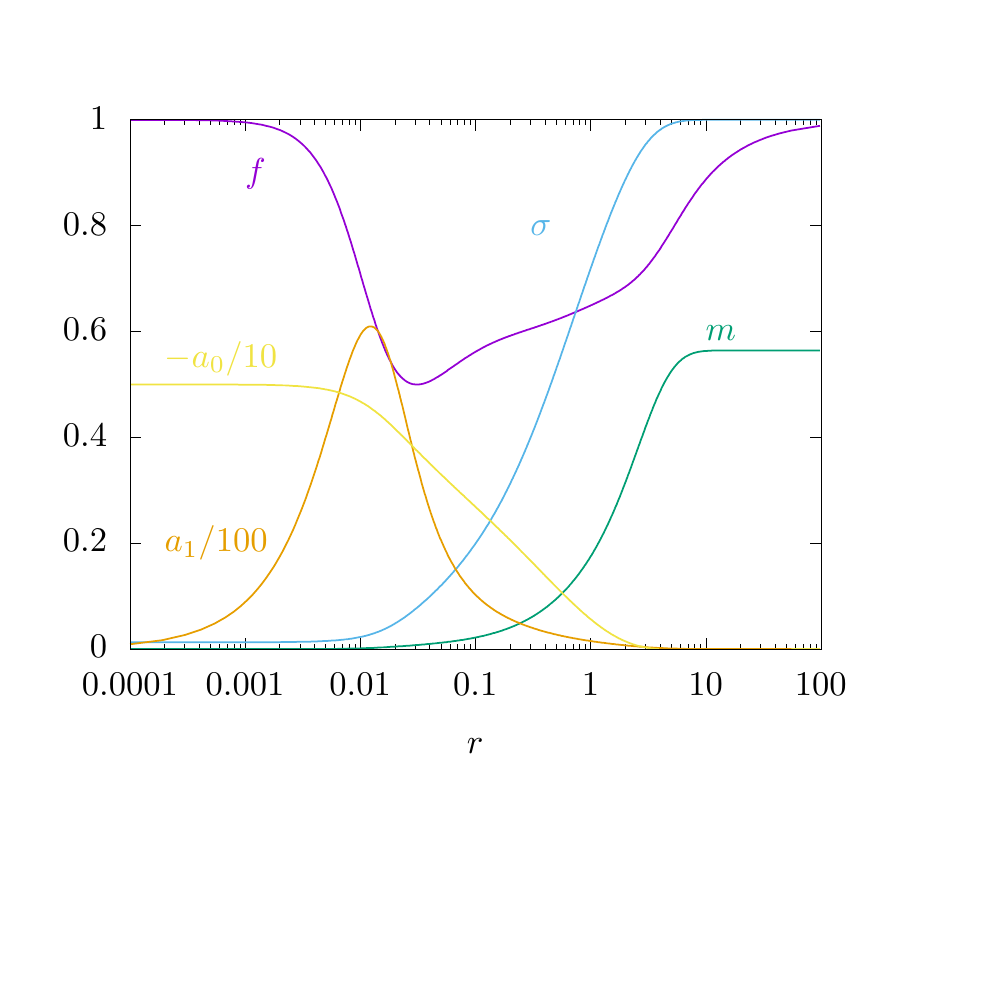}
\end{center}
\vspace{-2cm}
\caption{{\it (Left)}: The mass $M$, the charge $Q$ and the value of $\sigma(0)$ versus $\omega$ for solutions with $\gamma=0$ ($\kappa=0.1$, $\mu=1$, $\lambda=0$) corresponding to standard Proca stars. 
{\it (Right)}: 
An almost limiting solution of the Proca star branch with $a_0(0)=-5$. }
\label{fig:proca_gamma0}
\end{figure}

For $\gamma=0$, the Horndeski coupling term is switched off and the theory reduces to General Relativity with a complex scalar field.
The field equations then give rise to the
well-known Proca stars.
Their properties have been first discussed in \cite{Brito:2015pxa}, employing only a mass term for the vector field. 
In order to check the validity of our numerical procedure and be able to compare our results for the non-minimal model to that of minimally coupled gravity, we have re-constructed these solutions.
Our results are shown in Fig.~\ref{fig:proca_gamma0} (left), where we give the value of the mass $M$, the Noether charge $Q$ and the value of
the metric function $\sigma(r)$ at the origin, $\sigma(0)$, in dependence of $\omega$. 
Starting from $\omega_{\rm max}=\mu$, several branches of solutions exist (we have been able to construct three) that show the typical spiral-like behaviour. 
In the limiting solution $\sigma(0)\rightarrow 0$, while $f(r)$ remains perfectly well-behaved on the full interval $r\in [0:\infty)$, see Fig.~\ref{fig:proca_gamma0} (right) for the metric functions $f(r)$ and $\sigma(r)$ as well as the vector field functions $a_0(r)$ and $a_1(r)$ for
a solution with $a_0(0)=-5$. (Note that for $\gamma=0$, the system is symmetric under the exchange $(a_0,a_1) \rightarrow (-a_0,-a_1)$. )
This strongly suggests that the successive branches of Proca stars approach a configuration with a curvature singularity at the origin. 
In particular the Ricci scalar tends to infinity with
$R(0) \sim -6 \left(f''(0) + \frac{\sigma''(0)}{\sigma(0)}\right) + {\cal O}(r^2)$. 

The effect of a quartic self-interaction on the properties of Proca stars has been studied in \cite{Brihaye:2017inn}.

\subsection{$\gamma\neq 0$} 

In order to study the effect of the non-minimal coupling on the Proca star solutions discussed above, 
we now vary the coupling constant $\gamma$.
As in the previous case, 
we set $\kappa=0.1$, $\mu=1$, and $\lambda=0$. 
Since the sign of $\gamma$ plays a crucial role for 
the properties of the Proca stars,
we will now discuss the case of positive and negative $\gamma$ 
separately. 

\subsubsection{$\gamma > 0$}

\begin{figure}[h!]
\begin{center}
\includegraphics[width=.45\textwidth, angle =0]{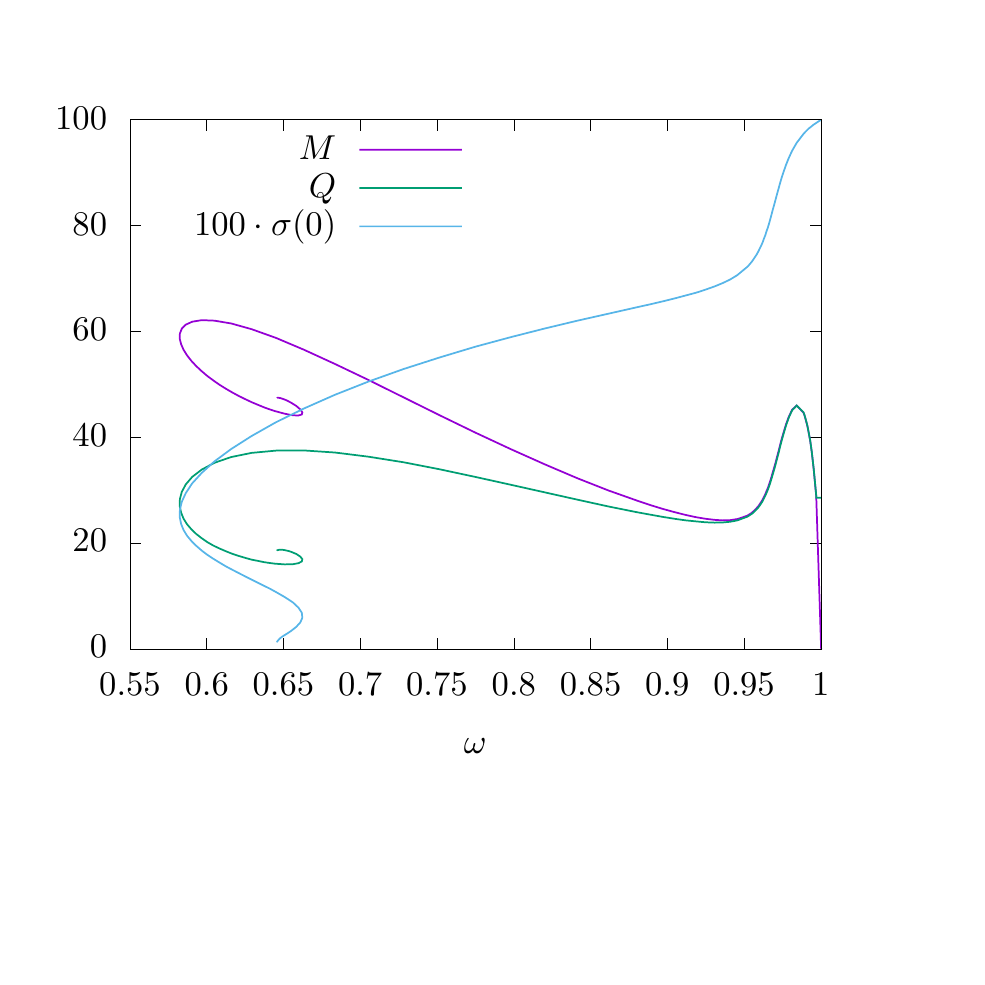}
\includegraphics[width=.45\textwidth, angle =0]{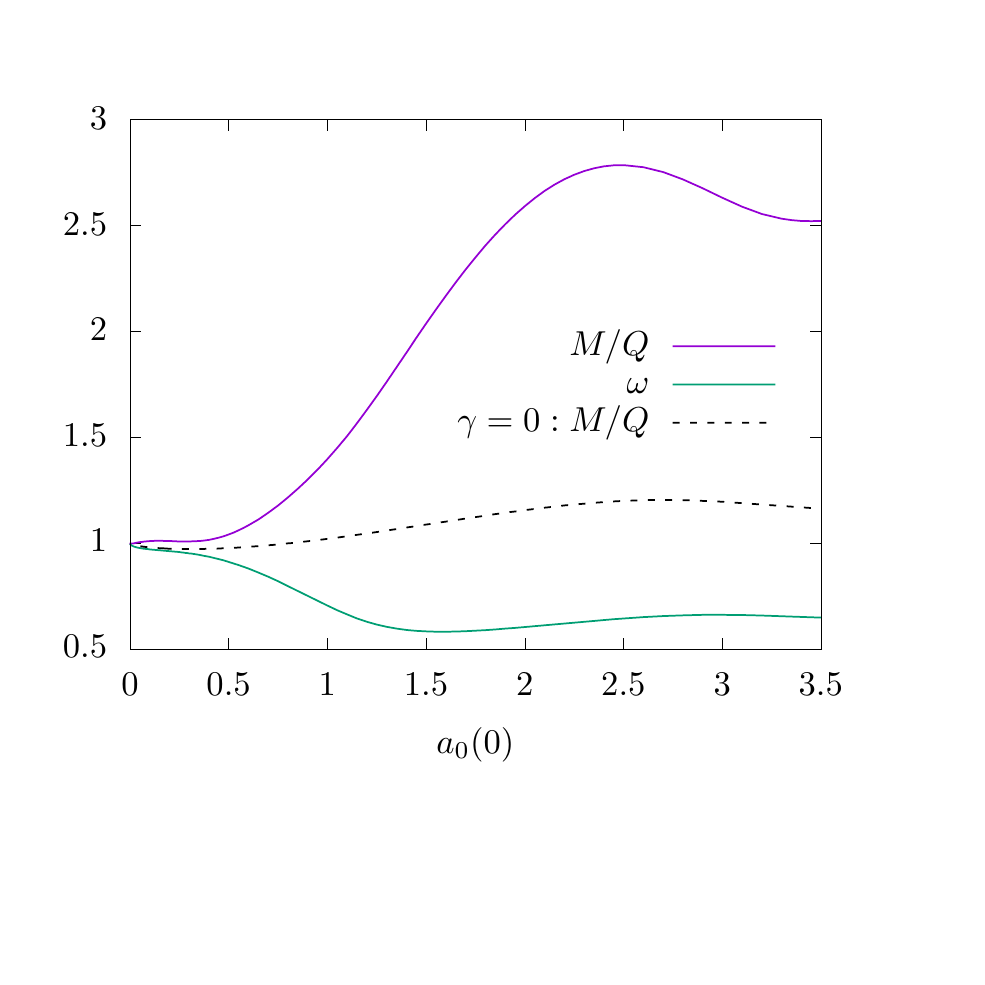}
\end{center}
\vspace{-2cm}
\caption{The mass $M$, the charge $Q$ and the value of $\sigma(0)$ versus $\omega$ (left) as well as the mass-to-charge ratio $M/Q$ and the value of $\omega$ versus $a_0(0)$ (right) for $\gamma=4.0$. For comparison, we also show the mass to charge ratio $M/Q$ in function of $a_0(0)$ for the $\gamma=0$ solutions (black dashed, right).}
\label{fig:proca_gamma_4}
\end{figure}

We find that for positive $\gamma$, 
important qualitative features found for the standard Proca stars are unaltered: we observe a spiraling 
behaviour and several branches of solutions when increasing the parameter $a_0(0)$. 
This is seen in Fig.~\ref{fig:proca_gamma_4} (left), where we show
the mass $M$, the Noether charge $Q$ and $\sigma(0)$ as functions of $\omega$ for $\gamma=4$. 
Compare this to Fig.~\ref{fig:proca_gamma_4}. 
We also show the mass-to-charge ratio $M/Q$ as well as the value of $\omega$ versus $a_0(0)$, see
Fig.~\ref{fig:proca_gamma_4} (right). 
Very similar to what has been observed in the $\gamma=0$ case, 
$a_0(0)$ can be increased to a maximal value, where $\sigma(0) \rightarrow 0$. 
However, for $\gamma=4$ we note that $M/Q > 1$ always.
Since $M/Q=1$ indicates the transition to an unbound system of $Q$ bosons of mass $\mu=1$, this means that
in this case the Proca stars should be 
unstable to decay into a system of free particles. 
This is of course different from the standard Proca case, where $M/Q<1$ along a large part of the fundamental branch, prohibiting such decay.
However, for sufficiently small positive values of 
$\gamma$ it is clear by continuity from the standard case
that $M/Q<1$ will still be realized along a part of the
fundamental branch, and thus stable solutions will be present.
Furthermore we observe that the non-minimal coupling allows for Proca stars with smaller values of $\omega$ as compared to the standard case.

\begin{figure}[h!]
\begin{center}
\includegraphics[width=.45\textwidth, angle =0]{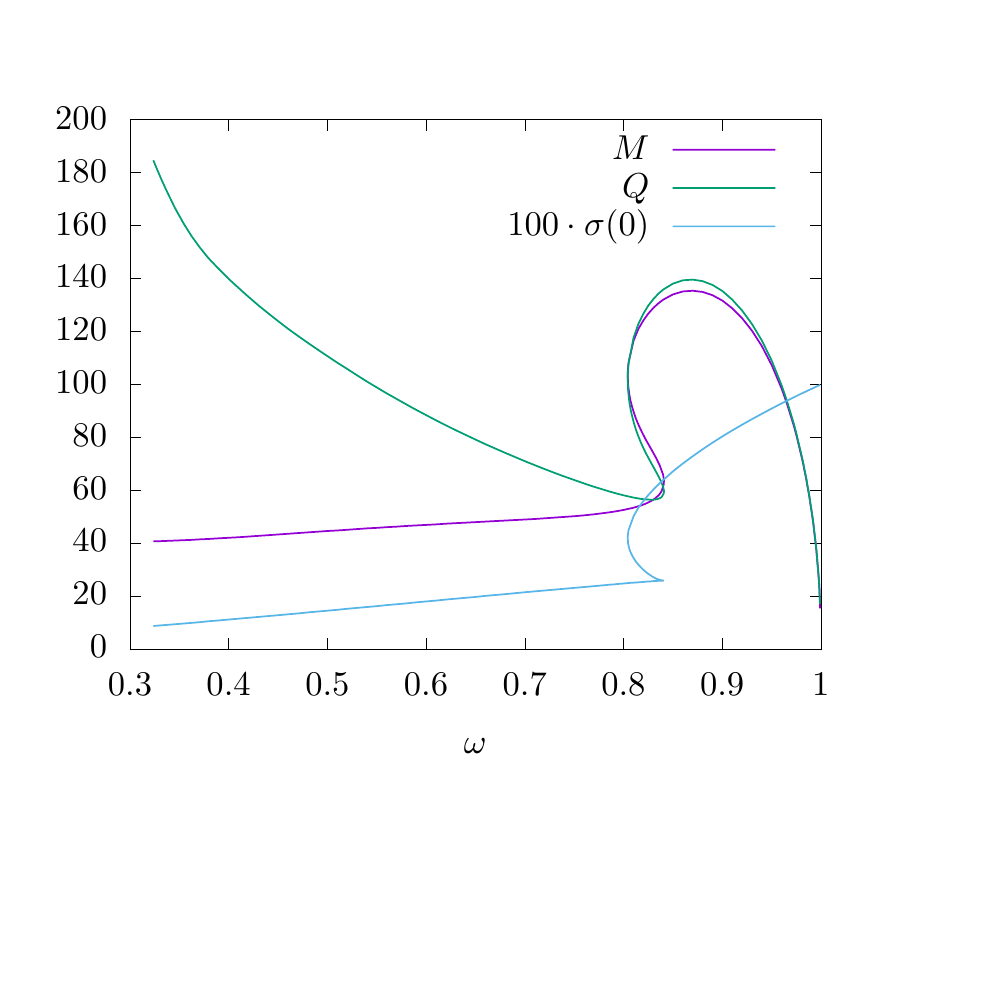}
\includegraphics[width=.45\textwidth, angle =0]{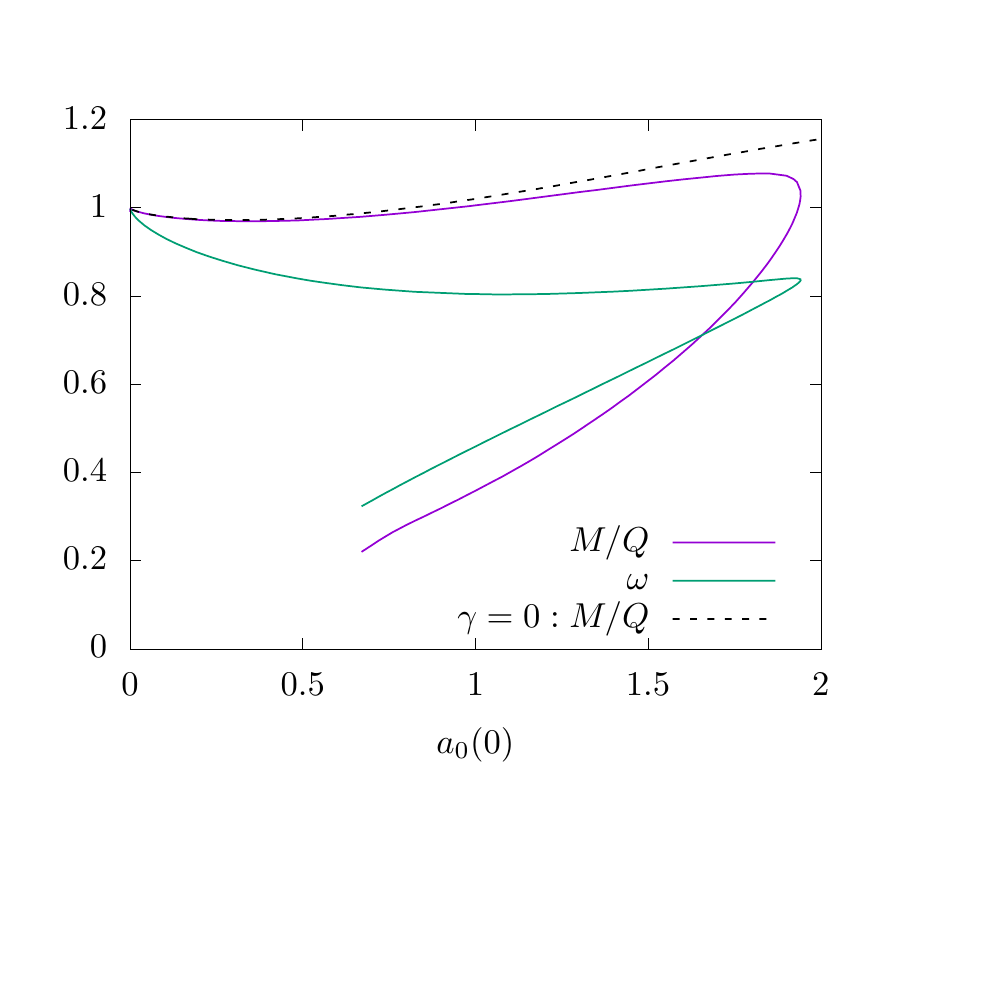}\\
\vspace{-2cm}
\includegraphics[width=.45\textwidth, angle =0]{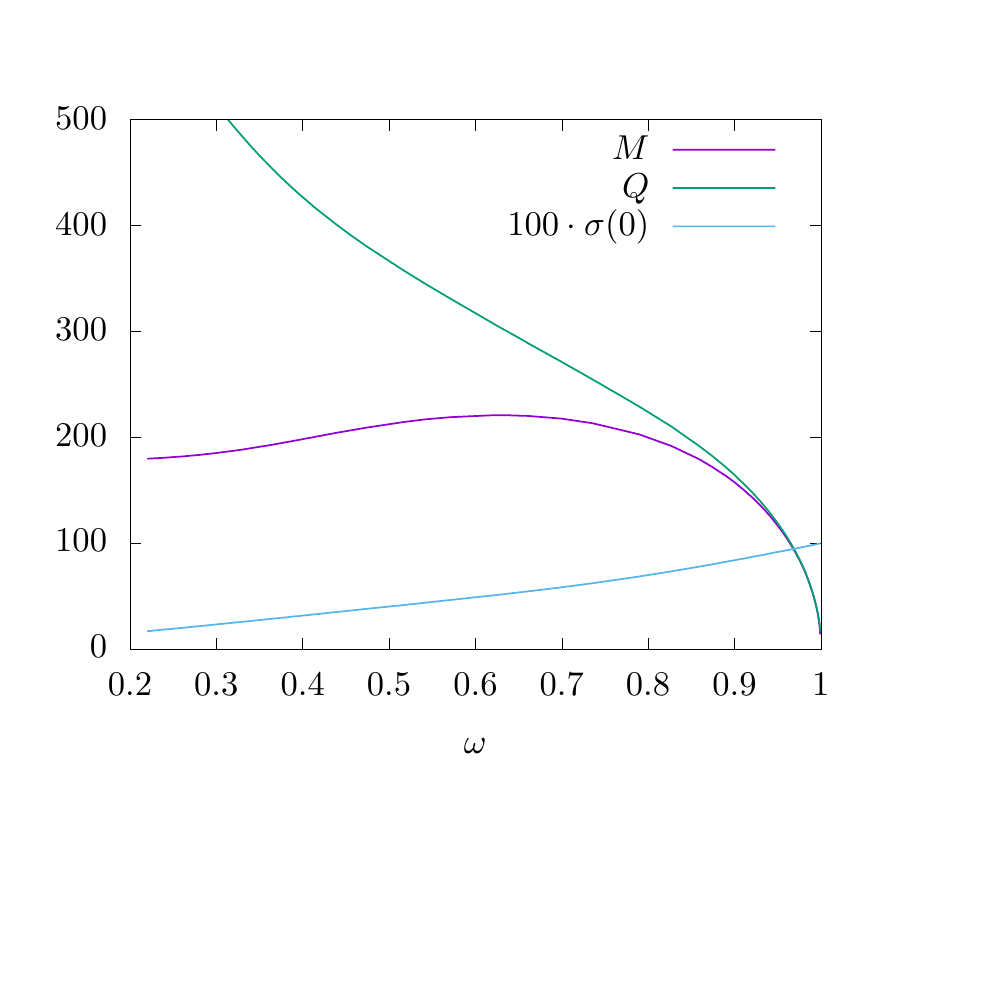}
\includegraphics[width=.45\textwidth, angle =0]{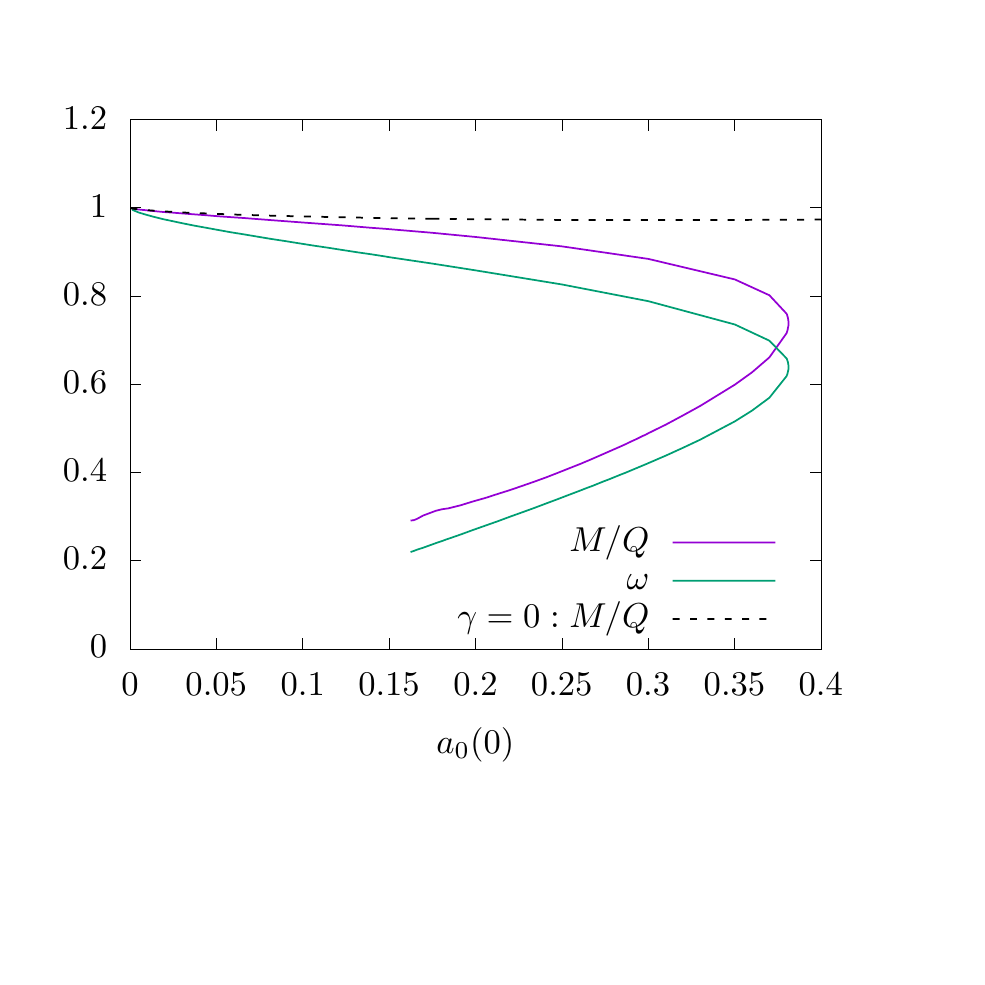}
\end{center}
\vspace{-2cm}
\caption{Same as Fig.\ref{fig:proca_gamma_4}, but for
$\gamma=-0.02$ (top left and right) and $\gamma=-0.4$ (bottom left and right), respectively. For comparison, we also show the mass to charge ratio $M/Q$ in function of $a_0(0)$ for the $\gamma=0$ solutions (black dashed, right).}
\label{fig:proca_gamma_negative}
\end{figure}

\subsubsection{$\gamma < 0$}

We now turn to negative values of the coupling constant $\gamma$.
Our results for two negative values of $\gamma$ are shown in 
Fig.~\ref{fig:proca_gamma_negative}.
As for positive $\gamma$, we observe that the non-minimal coupling allows
for lower values of $\omega$ also for negative $\gamma$.
However, as compared to the standard Proca case,
and the case of positive $\gamma$,
important qualitative features of the solutions have changed for negative $\gamma$. 
The main difference is that for sufficiently negative $\gamma$ (see Fig. \ref{fig:proca_gamma_negative} (bottom) for $\gamma=-0.4$) there is now a unique solution for a given value of $\omega$, i.e., we do not find a spiraling behaviour. Also, for $\gamma=-0.4$  the mass-to-charge ratio $M/Q < 1$ always. 
Thus the solutions should not be able to decay into $Q$ free particles.

To better explain the transition from positive to negative values of $\gamma$, we also present results for $\gamma=-0.02$, see Fig. \ref{fig:proca_gamma_negative} (top). We observe that the spiralling behaviour
close to the minimal value of $\omega$ has disappeared, but that the local maximum
of the mass $M$ and the charge $Q$ close to $\omega=1$ is still present. Moreover, as an intermediate case between the positive and negative value cases discussed above, there exist Proca stars with $M/Q < 1$ as well as with $M/Q \ge 1$ for $\gamma=-0.02$. Hence, depending on the choice of the value of $a_0(0)$ we would expect the Proca stars to be stable, resp. unstable, with respect to the decay into individual particles.

Additionally, the approach to criticality is very different as compared to the $\gamma \geq 0$ case.
When increasing $a_0(0)$ from zero, we find that a new phenomenon arises, when the maximal value of $a_0(0)$ is approached. 
The set of solutions no longer ends at this maximal value $a_{0,{\rm max}}(0)$. 
Instead, it can be continued by decreasing
$a_0(0)$ again. 
Thus at the maximal value two branches bifurcate smoothly and end.
For the data shown in Fig.~\ref{fig:proca_gamma_negative}, we find that $a_{0,{\rm max}}(0)\approx 0.38$ with a value of $\omega\approx 0.64$ for $\gamma=-0.4$, while  $a_{0,{\rm max}}(0)\approx 1.94$ with a value of $\omega\approx 0.84$ for $\gamma=-0.02$, respectively. This indicates that the larger the absolute value of $\gamma$, the smaller are $a_{0,{\rm max}}(0)$ and the corresponding $\omega$.

The new second branch of solutions exists for $a_0(0)\in [a_{0,{\rm cr}}(0):a_{0,{\rm max}}(0)]$,
with $a_{0,{\rm cr}}(0) < a_{0,{\rm max}}(0)$. 
As the critical value $a_{0,{\rm cr}}(0)$ is approached the metric function $f(r)$ 
tends to zero at some non-zero, intermediate value of the radial variable, $r=r_{\rm cr}$. 

\begin{figure}[h!]
\begin{center}
\includegraphics[width=.45\textwidth, angle =0]{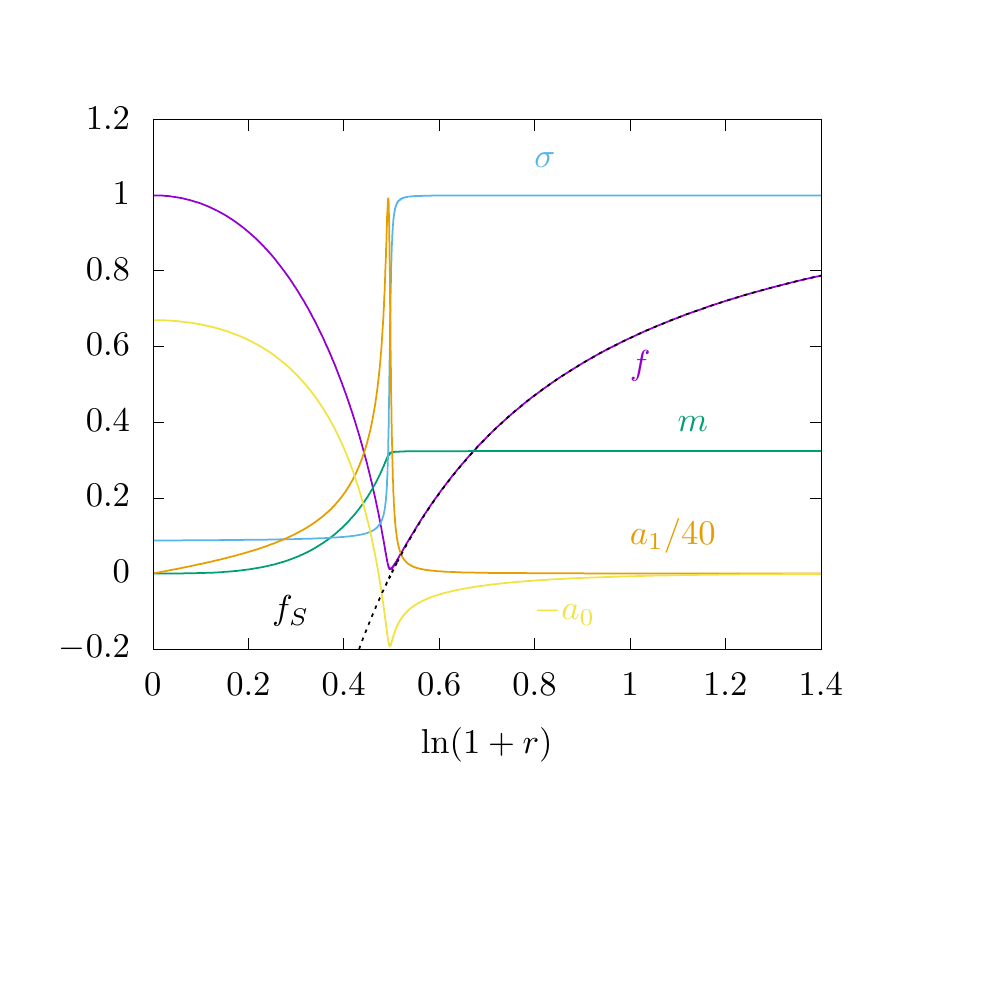}
\includegraphics[width=.45\textwidth, angle =0]{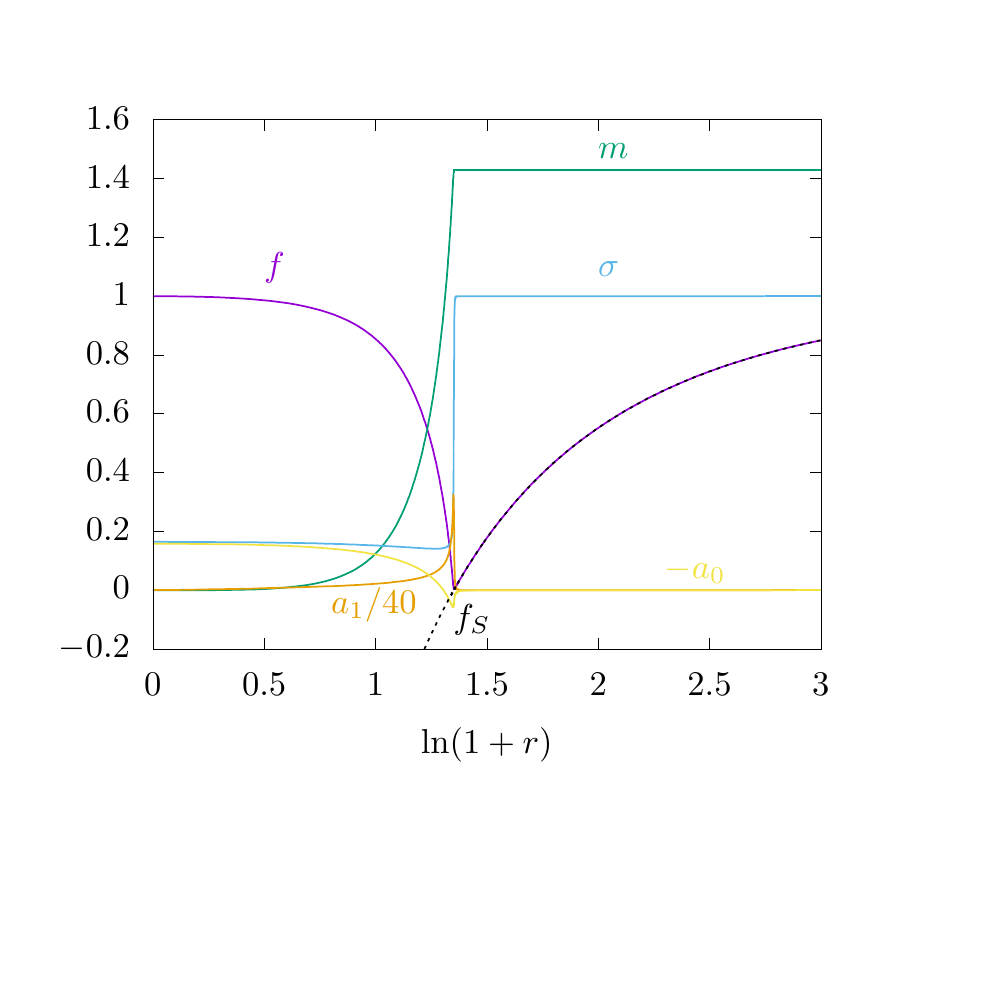}
\end{center}
\vspace{-2cm}
\caption{The profiles of the metric functions $f(r)$, $m(r)$, $\sigma(r)$ and the vector field functions
$a_1(r)$, $a_0(r)$
of a solution close to the limiting solution for $\gamma = -0.02$ corresponding to $\omega\approx 0.32$ (left) and for $\gamma = -0.4$ corresponding to $\omega\approx 0.22$ (right), respectively. For comparison we also show the metric function of the corresponding Schwarzschild solution $f_S=1-\frac{2M}{r}$ (black dashed) with $M=m(r\rightarrow \infty)$.} 
\label{fig:profile_critical}
\end{figure}

The profiles of the metric and vector field functions of a solution very close to the critical limit are shown in Fig.~\ref{fig:profile_critical} for
$\gamma=-0.02$ (left) and $\gamma=-0.4$ (right).
For comparison the figure also shows the metric function $f_{S}(r)$ for a Schwarzschild black hole with horizon radius $r_{\rm H}=r_{\rm cr}$ (black dashed line).
Intriguingly, there is almost perfect agreement 
of $f_{S}(r)$ with the function $f(r)$ for 
$r\ge r_{\rm cr}$, i.e., in the exterior region of the Schwarzschild black hole. 
Moreover,
the second metric function $\sigma(r)\equiv 1$ 
for $r > r_{\rm cr}$, which is again in agreement
with the metric of the Schwarzschild solution
since $\sigma_S(r)\equiv 1.$
This suggests that for $r > r_{\rm cr}$ 
the solution corresponds to 
a Schwarzschild black hole solution. 
Of course, a Schwarzschild black hole is a vacuum solution, so there should not be any non-trivial
matter fields in the exterior region $r > r_{\rm cr}$.
Inspection of the vector field function in
Fig.~\ref{fig:profile_critical} confirms,
that the vector field indeed vanishes in the
exterior region $r > r_{\rm cr}$.

In the interior region $r < r_{\rm cr}$, however,
the limiting solution features a finite vector field,
and also the metric functions differ strongly from the Schwarzschild metric functions.
Thus the spacetime of the limiting solution
is composed of an interior part with matter fields
and an exterior vacuum part, that are joined at the
horizon $r_{\rm H}=r_{\rm cr}$ of the exterior Schwarzschild
black hole, where the metric function $f(r)$ exhibits
a cusp, while the metric function $\sigma(r)$ exhibits a finite jump. 
Such jumps arise also for the two vector field functions.
In the standard case such a limiting behavior is not present.  Therefore it represents a new feature that appears for non-minimal coupling with sufficiently negative $\gamma$. 

\begin{figure}[h!]
\begin{center}
\includegraphics[width=.45\textwidth, angle =0]{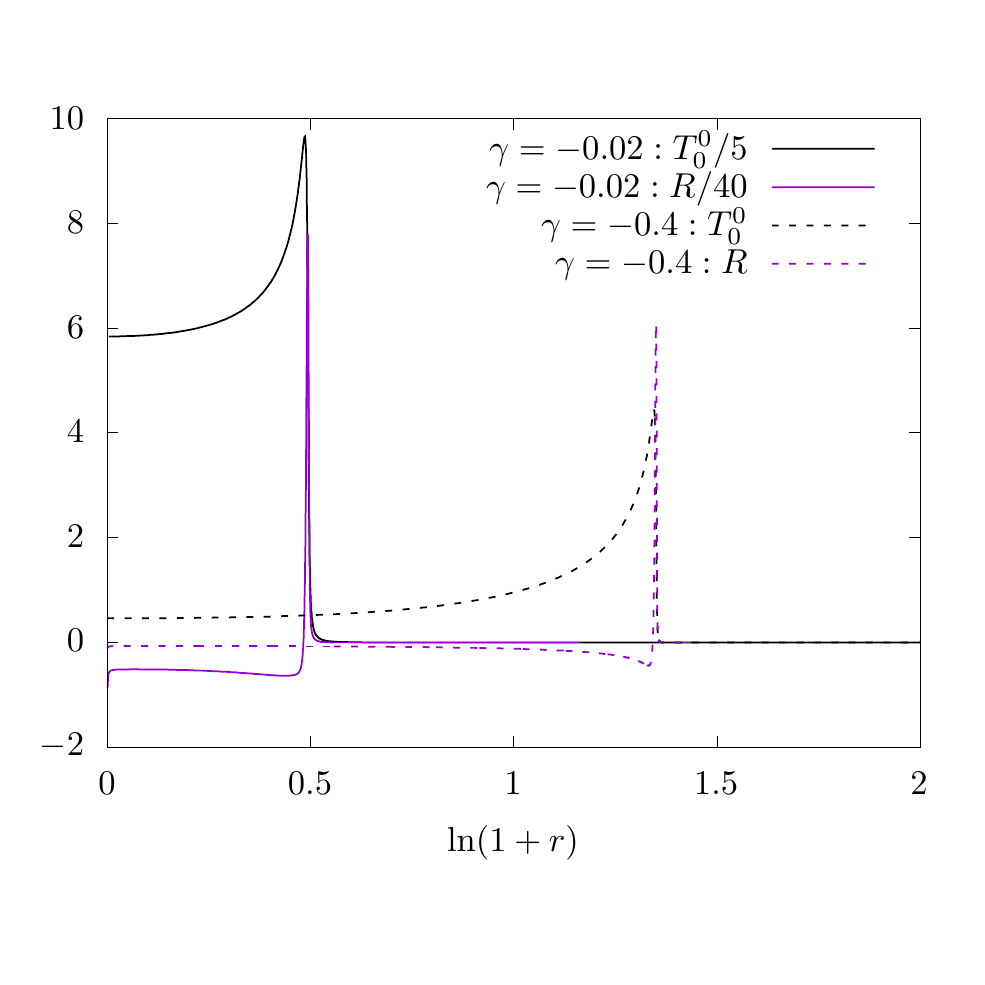}
\end{center}
\vspace{-1cm}
\caption{The $T_0^0$ component of the energy-momentum tensor (black) and the Ricci scalar $R$ (purple)  for the near criticial Proca star solutions
for $\gamma=-0.02$ (solid)  and $\gamma=-0.4$ (dashed), respectively.}
\label{fig:new1}
\end{figure}

In Fig.\ref{fig:new1}, we show the $T_0^0$ component of the energy-momentum tensor for the two near-critical solutions
in the case $\gamma=-0.02$ and $\gamma=-0.4$, respectively. Clearly, both $T_0^0$ and $R$ show a peak close to $r_{\rm H}=r_{\rm cr}$.

We remark that we have not encountered an analogous
limiting behavior before, where the spacetime splits
into an interior part with non-trivial matter fields
and an exterior Schwarzschild part. 
In contrast, the split into an interior part with non-trivial matter fields
and an exterior extremal Reissner-Nordstr\"om part
has been seen in many different circumstances,
ranging from gravitating monopoles \cite{Lee:1991vy,Breitenlohner:1991aa} 
(see also \cite{Volkov:1998cc})
to scalarized black holes \cite{Brihaye:2020yuv,Blazquez-Salcedo:2020crd}.
In that case the metric function $f(r)$ develops a
degenerate zero at $r_{\rm H}=r_{\rm cr}$.
In order for such a scenario to be able to take place, however, the Reissner-Nordstr\"om black hole must be a solution of the field equations. 
In the present case this is inhibited by the mass term of the vector field. 
The Schwarzschild solution
is, however, a solution of the field equations.

\begin{figure}[h!]
\begin{center}
\includegraphics[width=.45\textwidth, angle =0]{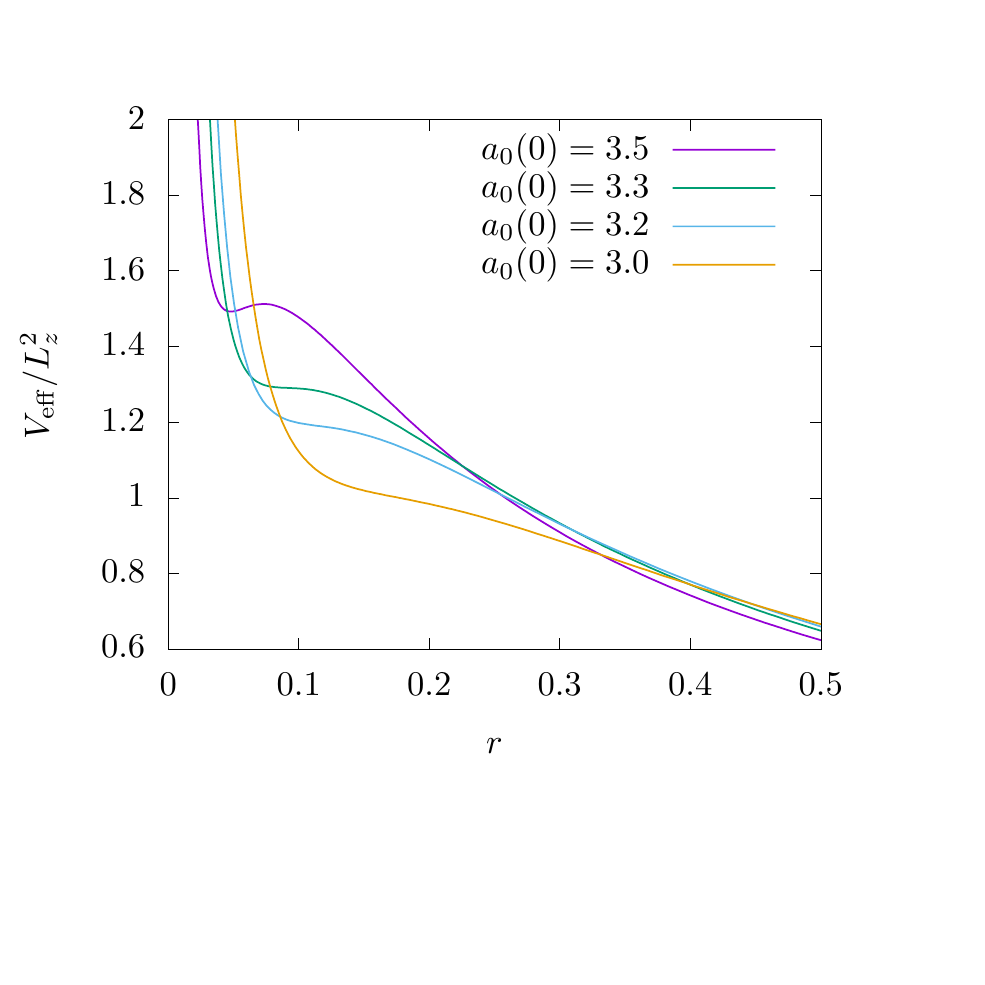}
\includegraphics[width=.45\textwidth, angle =0]{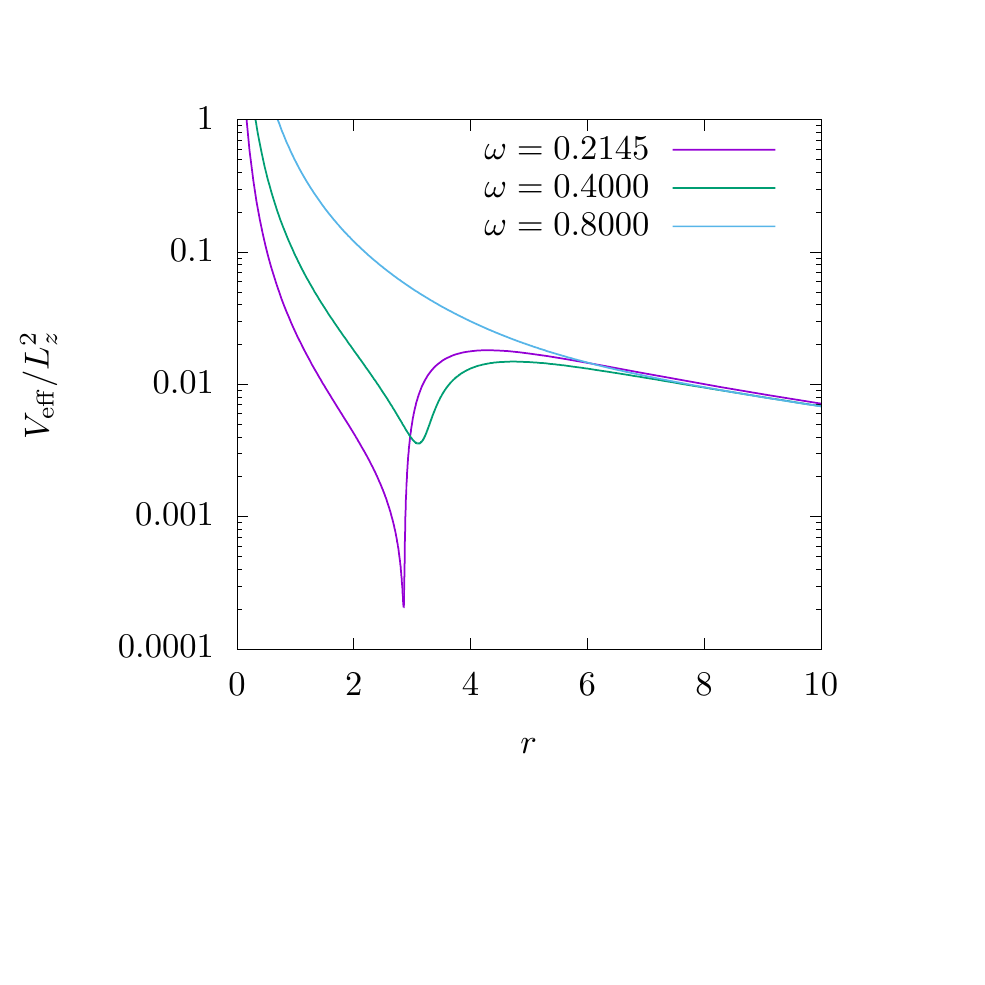}
\end{center}
\vspace{-2cm}
\caption{The effective potential  $V_{\rm eff}(r)/L_z^2$ (see (\ref{eq:geodesic})) for photons ($\varepsilon=0$) in the space-time of Proca star solutions
for $\gamma=4.0$ (left)  and $\gamma=-0.4$ (right), respectively. The values of $a_0(0)=3.0$, $3.2$, $3.3$, $3.5$ 
(left) correspond to $\omega\approx 0.662$, $0.658$, $0.655$, $0.649$, while the values of $\omega=0.2145$, $0.4000$, $0.8000$ (right) correspond to $a_0(0)\approx 0.158$, $0.287$, $0.286$. }
\label{fig:new2}
\end{figure}

It is then interesting to compare the compactness of these solutions with the Schwarzschild case. Since our solutions are certainly
not compact in the sense that they have a well defined radius outside which the energy density is strictly zero, we can compute the radius
$R_{99}$ (and $R_{95}$) 
of the sphere that contains $99\%$ (respectively $95\%$) of the mass $M$ of the Proca star solution. The values are given in Table \ref{table1}.
As expected, these values are very close to the Schwarzschild radius $r_{\rm S}=2M$ of these solutions. We have further investigated the 
effective potential appearing in the geodesic equation for test particles in this space-time. The geodesic equation can be written as
\begin{equation}
\label{eq:geodesic}
\sigma^2\dot{r}^2 + V_{\rm eff}(r)= E^2 \ \ , \ \     
V_{\rm eff}(r)=
f\sigma^2 \left(\frac{L_z^2}{r^2} - \varepsilon\right) \ , 
\end{equation}
where the dot denotes the derivative with respect to an affine parameter. Moreover, $\varepsilon$ takes on the value $0$ for massless particles and 
$-1$ for massive particles, respectively. 
We show the effective potential $V_{\rm eff}(r)/L_z^2$ of photons ($\varepsilon=0$) in the space-time of Proca stars for $\gamma=4.0$ and $\gamma=-0.4$, respectively, in Fig. \ref{fig:new2}. For $\gamma=4.0$ we find that the effective
potential possesses a positive-valued local maximum and a positive-valued local minimum for sufficiently
large values of $a_0(0)$. For $a_0(0)=3.5$ we find that the local minimum of the effective potential is located at
$r_{\rm V, min}\approx 0.048$ (see also Table \ref{table1}) and has value $V_{\rm eff}/L_z^2\approx 1.493$, i.e. a photon with $E^2/L_z^2=1.493$
wound move on a stable circular photon orbit with radius $0.048$ around the Proca star. Decreasing $\gamma$, we find that the location of this minimum moves to larger values of $r$ - see Table \ref{table1}. At values of $\gamma$ for which we
observe the phenomenon described above (see the plots for $\gamma=-0.4$ in Fig. \ref{fig:new2}), the location of the minimum of the effective potential is located roughly at the event horizon of the corresponding Schwarzschild black hole.
The space-time of the Proca star for $\gamma=-0.4$, $\omega=0.2145$ has $r_{\rm V, min}\approx 2.853$, while the mass
of the star is $M\approx 1.429$ corresponding to a Schwarzschild radius of $r_H=2.858$. 
The radius of the corresponding unstable photon orbit is $r_{\rm V, max}\approx 0.073$  for $\gamma=4.0$, $a_0(0)=3.5$.

\begin{table}[h!]
\centering
\begin{tabular}{|c|c|c|c|c|c|}
    \hline
    $\gamma$ & $\omega$  &  $a_0(0)$ & $r_{\rm V,min}$  \\
    \hline \hline
   $ -0.40$ & $0.2145$   & $0.158$   & $2.853$ &   \\
   \hline
    $-0.02$ & $0.8330$  & $1.900$ &   $0.635$ &    \\  
    \hline
      $0.00$ & $0.8900$  & $5.000$ &   $0.017$ &     \\  
    \hline
   $ 4.00$ & $0.6490$   & $3.500$   & $0.048$  \\
   \hline      
\end{tabular}
\caption{The location of the local minimum $r_{\rm V,min}$ of the effective potential (see also Fig. \ref{fig:new2}) for
some values of $\gamma$ and $\omega$ (or $a_0(0)$).}
\label{table1}
\end{table}

\begin{figure}[h!]
\begin{center}
\includegraphics[width=.45\textwidth, angle =0]{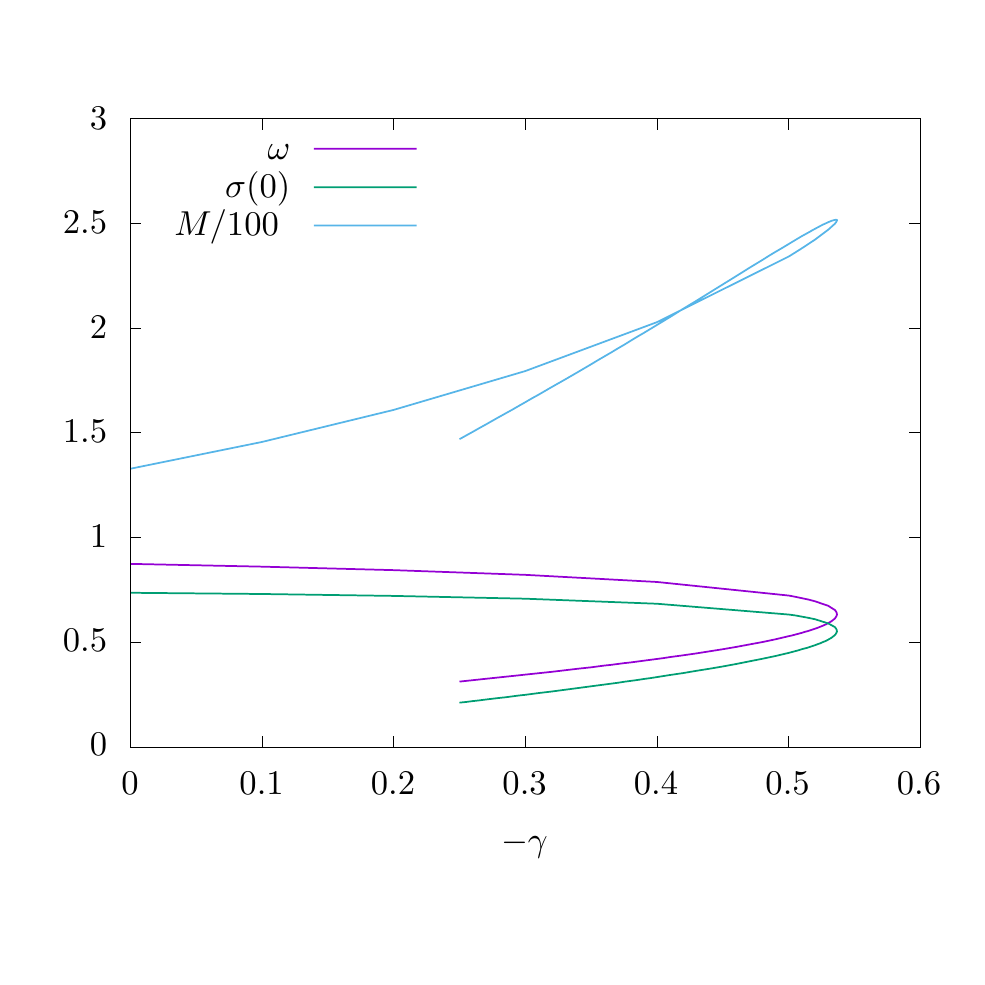}
\end{center}
\vspace{-1cm}
\caption{The value of $\omega$, of $\sigma(0)$ as well as of the mass $M$ versus the coupling constant $-\gamma$ for fixed $a_0(0)=-0.3$.
 }
\label{fig:data_a_0_fixed}
\end{figure}

We have also constructed branches of solutions for fixed value of $a_0(0)$ and varying non-minimal coupling parameter $\gamma$. 
Our results for $a_0(0)=-0.3$ are shown in Fig.~\ref{fig:data_a_0_fixed}. 
Obviously, the solutions corresponding to $\gamma < 0$ exist only up to a maximal value of $\vert\gamma\vert$, where again two branches of solutions bifurcate smoothly and end.
These branches that can, for instance, be distinguished by their values of $\omega$. 
One of these branches connects to the standard Proca star in the limit $\gamma\rightarrow 0$.
When decreasing $\gamma$ from zero along this branch, we find that the mass increases and the frequency decreases until the minimal possible value of $\gamma$, $\gamma_{\rm min}$, is reached.
Here this branch merges smoothly with the second branch of solutions, that exists for $\gamma\in [\gamma_{\rm min}:\gamma_{\rm cr}]$, 
where $\gamma_{\rm min} < \gamma_{\rm cr}$. 
For $a_0(0)=-0.3$ we find that 
$\gamma_{\rm min}\approx  - 0.54$, 
while $\gamma_{\rm cr}\approx -0.25$. 
The limit $\gamma \to\gamma_{\rm cr}$ is analogous to the limit described above with a 
zero of the metric function $f(r)$ forming at some intermediate value of the radial variable, $r=r_{\rm cr}$. 

\section{Conclusion}

We have constructed Proca stars in a vector-tensor theory, obtained by a generalizing the vector-tensor theory of Horndeski \cite{Horndeski:1976gi} by promoting the vector field to a complex massive field.
In this theory the vector field is non-minimally coupled to gravity,
where the strength of the corresponding coupling term is regulated by a coupling constant $\gamma$.
For vanishing $\gamma$ General Relativity and the
standard Proca stars with their well-known features are recovered.

As the coupling constant is increased or decreased from zero the properties of the Proca stars start to change.
For positive values of $\gamma$ the spiraling pattern of the solutions is retained, but the global charges $M$ and $Q$ decrease.
Since the charge $Q$ decreases faster, the Proca star solutions no longer present bound systems for sufficiently large values of the coupling, since the mass-to-charge ratio is always greater than one.
Thus stability of the solutions should be lost, since their decay would be energetically favorable.

For negative values of $\gamma$, on the other hand, the changes with respect to the standard Proca case are even stronger, when the magnitude of $\gamma$ is sufficiently large. 
In this case no trace of the spiraling pattern of the standard Proca stars is left.
Instead the solutions can be continued to much smaller values of $\omega$ before they cease to exist. 
The mass shows a single maximum, while the charge continues to rise monotonically. 
Consequently, the mass-to-charge ratio of these configurations is always smaller than one, with the bosons getting continuously stronger bound, as the limiting configuration is approached.

The limiting configuration in the case of (sufficiently) negative coupling possesses rather surprising features. 
It consists of two distinct parts, an interior part with matter fields and an exterior vacuum part.
Both parts are joined at a critical radius $r_{\rm cr}$, where the exterior Schwarzschild solution features its event horizon. 
The transition at $r_{\rm cr}$ is, however, not smooth. 
This limiting configuration is vaguely reminiscent of the limiting configurations occurring in certain non-Abelian or scalarized solutions.
However, in those cases the exterior solution corresponds to an extremal Reissner-Nordstr\"om solution with a degenerate horizon. 
In the present case this would not be possible, since only a Schwarzschild black hole but not a Reissner-Nordstr\"om black hole is a solution of the field equations. 

Interesting future work in this vector-tensor theory will be the inclusion of rotation to generate rotating generalized Proca stars. 
It will then be tempting to subsequently insert a horizon.
In this latter case generalized Kerr black holes with Proca hair will result, where the frequency of the Proca field will be synchronized with the rotational velocity of the event horizon  \cite{Herdeiro:2016tmi}.

\section*{Acknowledgement}

BH, BK and JK gratefully acknowledge support by the
DFG Research Training Group 1620 {\sl Models of Gravity}
and the COST Actions CA15117 {\sl CANTATA} 
and CA16104 {\sl GWverse}.

\end{document}